\documentclass[12pt]{article}

\usepackage{amsmath}
\usepackage{amsfonts}
\usepackage{amssymb}

\usepackage{graphicx}
\usepackage{subfigure}

\begin{document}
%
\title{D-optimal design of b-values for accurate intra-voxel incoherent motion imaging}
%
%
%

\author{Mario~Sansone,
        ~Roberta~Fusco
        ~and~Antonella~Petrillo
\thanks{M. Sansone is with the Department
of Electrical Engineering and Information Technologies, University `Federico II', Naples,
Italy, e-mail: msansone@unina.it}
\thanks{R. Fusco and A. Petrillo are with National Cancer Institute `Fondazione Pascale', Naples Italy.}
}

%
%

\markboth{Journal of \LaTeX\ Class Files,~Vol.~6, No.~1, January~2007}%
{Shell \MakeLowercase{\textit{et al.}}: Optimal design for IVIM}
%



\maketitle

\begin{abstract}
The aim of this paper is to optimally design the set of $b$-values
for diffusion weighted MRI 
with the aim of accurate estimation
of intra-voxel incoherent motion (IVIM) parameters
($f$ perfusion fraction, $D_s$ slow diffusion, $D_f$ fast diffusion)
according to the model developed by Le Bihan.

Previous studies have addressed the design in a Monte Carlo fashion.
Due to huge computation times,
 this approach is feasible only for a limited number of values of the parameters (local design):
 however, as the parameters of a specific exam are not known \emph{a priori}, it would be desirable to optimise $b$-values over a \emph{region} 
 of parameters.

In order to overcome this issue, we propose to use a D-optimal design approach.
The optimal combination of $b$-values
can be chosen from a candidate set of predefined values
taken from the literature. 
Our study has two key results: first, the optimal design 
does not depend on perfusion fraction: this allow to perform a search over a 2D parameter space instead of 3D; 
second, as an exhaustive search over all possible designs would  still be 
 time consuming, we proposed an algorithm to find  
an approximate solution can be found 
very quickly. 
\end{abstract}


\section{Introduction}
%
%
%
%
Optimal  choice of $b$-values for
acquisition of  diffusion weighted MR (DW-MR) data
is still under debate \cite{Koh:2011aa,zhang2013cramer,lemke2011toward,jambor2014optimization}.
In particular, optimal $b$-values should allow
for accurate  estimation of IntraVoxel Incoherent Motion (IVIM) parameters
such as diffusion coefficients and perfusion fraction
\cite{le1988separation}.

In principle, for accurate estimation of IVIM parameters,
as many $b$-values as possible
should be used.
However, this is in contrast with 
clinical requirements such as the duration
of the diffusion exam, 
the confort of the patient etc...
Therefore, in clinical literature \cite{lemke2011toward,jambor2014optimization}
typically up to 11 $b$-values are commonly used.
Depending on the specific field of view,
in typical examinations, the acquisition of one $b$-value
can take a few minutes to complete.
However, accurate estimation of IVIM parameters
can still be achieved even with a small number of $b$-values 
if they  have been opportunely chosen.

Previous attempts to design optimal $b$-values have been reported
in \cite{jambor2014optimization} and \cite{lemke2011toward}.
They tackled the problem using a Monte Carlo approach.
This approach is time consuming and has not been applied to 
find an optimal design for a large region within the parameter space
but only for the \emph{local} optimal design for a few values of the parameters.
However, it would be desirable to design optimal values for a large region 
because the specific values for each patient are not known \emph{a priori}, of course.

This issue might be overcome using a D-optimal approach.
To the best of our knowledge, 
the problem of $b$-values design
has not yet been addressed using a D-optimal approach.
This has a sounding mathematical basis
and can lead, as we show in this manuscript, to a fast  choice of optimal $b$-values 
(among a set of predefined values) over a region of the parameters space.

The specific aim of this paper is to show how the design of the $b$-values
can be simplified performing the search only in the space of the diffusion coefficients
and to propose a fast algorithm for finding an (approximate) design over an entire region of this space.
The optimal combination of $b$-values
has been chosen from a set of predefined values
taken from the literature.
The reliability of the approximate design has been evaluated
on the basis of the Cramer-Rao lower bound.

\section{Methods}

\subsection{IVIM modelling}
The most used model separating the contribute of diffusion and perfusion in
intra-voxel incoherent motion has been developed by Le Bihan \cite{le1988separation}.
According to this pioneering paper, the signal intensity of diffusion weighted MRI can be described by eq. \ref{eq:ivim1}:
\begin{eqnarray}\label{eq:ivim1}
& & S(b , S_0 ,f,D,D^*) = \\
\nonumber & = & S_0 \left[ (1-f) \exp(-b D) + f \exp\left(-b (D + D^* )\right) \right] 
\end{eqnarray}
where $b$ is a factor depending on the gradient pulse sequence, 
$D$ is the diffusion coefficient of water,
$D^*$ is a pseudo-diffusion coefficient describing blood microcirculation,
$f$ is the fraction of water flowing in perfused capillary,
$S_0$ is the signal intensity when $b=0$.

As suggested in \cite{jambor2014optimization}, eq. \ref{eq:ivim1} can be rearranged as in    eq. \ref{eq:ivim2}:
\begin{eqnarray}\label{eq:ivim2}
& & S(b,S_0,f,D_s,D_f) = \\
\nonumber & = & S_0 \left[ (1-f) \exp(-b D_s) + f \exp\left(-b D_f \right) \right]
\end{eqnarray}
where $D_s = D$ represents the \emph{slow} component of diffusion
and $D_f = D+D^*$ represents the \emph{fast} component of diffusion. 
This form slightly simplifies formula manipulation and will be used in the following.

Units of the diffusion coefficients are $\mathrm{mm}^2/\mathrm{s}$
while $b$ is measured in $\mathrm{s}/\mathrm{mm}^2$.
Typically \cite{le1988separation,jambor2014optimization}, 
 $D_f$ is about $20\cdot10^{-3} \mathrm{mm}^2/\mathrm{s}$
while  $D_s$ is about  $1\cdot10^{-3} \mathrm{mm}^2/\mathrm{s}$.
In general $D_f>>Ds$.
Values for $b$, which is the focus of this paper,
falls typically in the range $0 \rightarrow 1000 \quad \mathrm{s}/\mathrm{mm}^2$.
Perfusion fraction $f$ has no units and ranges $0<f<1$.

\subsection{Noise on diffusion weighted data}\label{sec:noise}
It is well known that, because of noise superimposed to receiving antennas, 
the measured signal intensity ($S_m$) of diffusion  MRI data
has a Rician distribution \cite{gudbjartsson1995rician} as in eq. \ref{eq:rician1}:
\begin{equation}\label{eq:rician1}
p(S_m;S,\sigma) = \frac{S_m}{\sigma^2} \exp \left( - \frac{S_m^2 + S^2}{2\sigma^2}\right) I_0\left(\frac{S_m S}{\sigma^2}\right)
\end{equation}
where $S$ is the  signal intensity without noise (which should be given by the IVIM model eq. \ref{eq:ivim2}),
$\sigma$ is the noise level,
$I_0$ is the modified Bessel function of the first kind.

It the limit of high Signal Noise Ratio (SNR, $S/\sigma > 3$) the distribution can be well approximated by a Gaussian
distribution \cite{gudbjartsson1995rician,kristoffersen2007optimal}:
\begin{equation}
p(S_m;S,\sigma) \approx \frac{1}{\sqrt{2 \pi \sigma^2}} \exp \left( -\frac{(S_m-\sqrt{S^2+\sigma^2})^2}{2 \sigma^2}\right)
\end{equation}

In our experience, on diffusion images of the prostate at several $b$-values
from 0 to 1000, the approximation $S >> \sigma$ is very well satisfied.
In fact, for voxels within region of interest (ROI) the measured intensity $S_m(b)$ is about $100$ 
for $b=0$
decreasing to about $50$ when $b=1000 \quad \mathrm{s}/\mathrm{mm}^2$;
the $\sigma$ parameter evaluated outside the field of view,
gives a value of about $2$ ($S/\sigma \approx 50$).
At our institution images have been obtained using a Siemens scanner, 1.5 T, 
with pulse sequences EP (SK/SP/OSP), TR  = 7500 ms, TE = 91 ms, 3 averages, 
flip angle = 90 deg.

Moreover, the SNR measures reported in \cite{jambor2014optimization}
confirm that the noise level $\sigma$ is typically very low 
with respect to the signal level $S$.

Generalising the previous considerations, in the following we will assume 
that the hypothesis of Gaussian approximation is satisfied;
additionally, we will assume that to a very good approximation $\sqrt{S^2 + \sigma^2} \approx S$ as $S >> \sigma$, 
and therefore the measured signal is:
\begin{equation}
S_m = S + \epsilon
\end{equation}
with additive noise $\epsilon \sim\mathcal{N}(0,\sigma)$.

\subsection{IVIM model for $b$ optimisation}
\label{sec:ivimModel4optim}
It is a common approach in IVIM literature to normalise DW data $S_m(b)$
dividing by $S_m(0)$ because in so doing the model simplifies to a three parameters model.
Typically, a Least Squares (LS) fitting is applied \cite{fusco2015use,le1988separation,jambor2014optimization}
assuming, implicitly \cite{bates1988nonlinear}, that the noise onto the normalised data $S_m(b)/S_m(0)$ is additive gaussian.

However, as it has been observed in the previous section, 
while the noise superimposed on $S$ can be well approximated (for $S/\sigma > 3$) by  a gaussian distribution,
the distribution of $S_m(b)/S_m(0)$ becomes a Cauchy distribution instead \cite{probability1984random}.
As a consequence, the LS approach to non linear fitting,
which is derived within the framework of  Maximum Likelihood 
under the hypothesis of spherical Gaussian distribution, 
might be not adequate in this case.

With this in mind, for $b$ design purposes (section \ref{sec:design}) and for evaluation of attainable accuracy (section \ref{sec:accuracy})
 we consider here a complete 4 parameters model $S(b,S_0,f,D_s,D_f)$ 
including $S_0$ among the parameters,
instead of the common approach to consider the normalised signal.

\subsection{D-optimal design}\label{sec:design}

As underlined in the previous section,
DW signal intensities in real cases 
can be be typically well approximated  by a Gaussian distribution, implying that the measured signal
$S_m(b)$ has an IVIM component plus an additive noise component as in eq. (\ref{eq:ivim3}):
\begin{equation}\label{eq:ivim3}
S_m(b;S_0,\boldsymbol{\theta}) = S(b,S_0,\boldsymbol{\theta}) + \epsilon 
\end{equation}
with $\epsilon$ having Gaussian distribution $\mathcal{N}(0,\sigma)$
with zero mean and standard deviation $\sigma$,
and $\boldsymbol{\theta} = [f,D_s,D_f]^T$. 

Under this hypothesis it is possible to use 
the theory for optimal design of experiments described, for example, in  \cite{bates1988nonlinear,fedorov1972theory}.
Optimal design is based upon the computation of the Fisher information matrix of the IVIM model,
which we perform in the following.

The Fisher information matrix corresponding to \emph{all} parameters $\mathbf{p} = [S_0, f, D_s, D_f]^T $ is given by eq \ref{eq:fisher}:
\begin{eqnarray}\label{eq:fisher}
\nonumber \mathbf{M} &=& \{ M_{ij} \}= \sigma^{-2} \sum_{k=1}^N \frac{\partial S(b_k)}{\partial p_i} \frac{\partial S(b_k)}{\partial p_j} =  \\
&=& \sigma^{-2} \frac{\partial S(\mathbf{b}^T)}{\partial \mathbf{p}} \frac{\partial S(\mathbf{b})}{\partial \mathbf{p}^T} 
\end{eqnarray}
where the partial derivatives of $S$ are evaluated at $b_k$ with $\mathbf{b} = [b_1,b_2,\ldots,b_N]^T$.
However, we note that the model in eq. (\ref{eq:ivim2})
is conditionally linear in the parameter $S_0$, thus it can be re-formulated as:
\begin{equation}\label{eq:reducedModel}
S(b;S_0,\boldsymbol{\theta}) = S_0 \cdot g(b,\boldsymbol{\theta})
\end{equation}
where 
\begin{equation}
g(b,\boldsymbol{\theta}) = (1-f) \exp(-b D_s) + f \exp\left(-b D_f \right)
\end{equation}
and therefore the Fisher Matrix can be rewritten as:
\begin{eqnarray}
\mathbf{M} & = & \sigma^{-2}
\left( \begin{array}{cc}
1 & \mathbf{0} \\
 \mathbf{0} & S_0 \mathbf{I} \\
\end{array} 
\right)
\mathbf{J}^T \mathbf{J}
\left( \begin{array}{cc}
1 & \mathbf{0} \\
\mathbf{0} & S_0 \mathbf{I} \\
\end{array} 
\right) \\
\nonumber & = & \sigma^{-2} \mathbf{A}^T(S_0) \mathbf{J}^T \mathbf{J A}(S_0)
\end{eqnarray}
where $\mathbf{I}$ is the identity matrix of size $3\times3$, $\mathbf{A}(S_0)$ depends only on $S_0$ and
\begin{eqnarray}\label{eq:jacobian}
\nonumber \mathbf{J}(\mathbf{b},\boldsymbol{\theta}) & = &  \left( \begin{matrix}
g(b_1,\boldsymbol{\theta}) & \frac{\partial g(b_1,\boldsymbol{\theta})}{\partial f} & \frac{\partial g(b_1,\boldsymbol{\theta})}{\partial D_s} & \frac{\partial g(b_1,\boldsymbol{\theta})}{\partial D_f}  \\ 
g(b_2,\boldsymbol{\theta}) & \frac{\partial g(b_2,\boldsymbol{\theta})}{\partial f} & \frac{\partial g(b_2,\boldsymbol{\theta})}{\partial D_s} & \frac{\partial g(b_2,\boldsymbol{\theta})}{\partial D_f}  \\ 
\vdots & \vdots & \vdots & \vdots \\ 
g(b_N,\boldsymbol{\theta}) & \frac{\partial g(b_N,\boldsymbol{\theta})}{\partial f} & \frac{\partial g(b_N,\boldsymbol{\theta})}{\partial D_s} & \frac{\partial g(b_N,\boldsymbol{\theta})}{\partial D_f}
\end{matrix} 
\right) \\
& = &  
[\mathbf{g},~\partial_f \mathbf{g},~\partial_{D_s} \mathbf{g},~\partial_{D_f} \mathbf{g}]
\end{eqnarray}
where the four columns of $\mathbf{J}$ have been indicated by vectors for simplicity.

Further simplification can be obtained using the following notation: 
\begin{eqnarray}
\nonumber \mathbf{E}_s &=& \exp( - \mathbf{b} D_s) \\
\mathbf{E}_f &=& \exp( - \mathbf{b} D_f) 
\end{eqnarray}
which gives $\mathbf{g}  = (1-f)\mathbf{E}_s + f \mathbf{E}_f$ and $\partial_f \mathbf{g} = \mathbf{E}_f - \mathbf{E}_s$.
Moreover, we indicate with $\circ$ the Hadamard product between two vectors e.g.: 
\begin{equation}
\mathbf{x} \circ \mathbf{y} = [x_1 y_1, \ldots, x_N y_N]^T;
\end{equation}
which gives $\partial_{D_s} \mathbf{g} = -(1-f)\mathbf{b} \circ \mathbf{E}_s$ and
$\partial_{D_f} \mathbf{g} = -f \mathbf{b} \circ \mathbf{E}_f$.

With this notations the matrix $\mathbf{J}$ can be rewritten:
\newcommand\scalemath[2]{\scalebox{#1}{\mbox{\ensuremath{\displaystyle #2}}}}
\begin{eqnarray}
\nonumber & &\mathbf{J}(\mathbf{b},\boldsymbol{\theta}) = \\ 
\nonumber & = &
 [\mathbf{E}_s, \mathbf{E}_f ,\mathbf{b} \circ \mathbf{E}_s, \mathbf{b} \circ \mathbf{E}_f ] 
\left( 
\scalemath{0.75}{
\begin{matrix}
1-f & -1 & 0 & 0 \\
f &  1 & 0 & 0 \\
0 & 0 & -(1-f) & 0 \\
0 & 0 & 0 & -f
\end{matrix}
}
\right) \\
& = &  \mathbf{H}(\mathbf{b},D_s,D_f) \mathbf{B}(f)
\end{eqnarray}

According to the  D-optimal design approach
we must find the $b$ values maximizing the determinant ($|| \cdot ||$) of the Fisher matrix:
\begin{eqnarray}
\nonumber & & ||\mathbf{M}(\mathbf{b},S_0,\boldsymbol{\theta})|| = \\
&=& \sigma^{-2} || \mathbf{A}(S_0) ||^2 || \mathbf{B}(f) ||^2 || \mathbf{H}^T \mathbf{H} (\mathbf{b},D_s,D_f) || 
\end{eqnarray}
it is clear that the optimal value for $\mathbf{b}$ can be found independently from $S_0$ and $f$:
\begin{equation}\label{eq:maxFisher}
\mathbf{b}_{opt} = \arg \max_{\mathbf{b} \in \mathcal{B}} \left( \min_{D_s, D_f \in\boldsymbol{\Theta}}\left\lVert \mathbf{H}^T \mathbf{H} \right\rVert \right) 
\end{equation}
where $\mathcal{B}$ is the \emph{design} space (the set of all candidate $b$-values) and $\boldsymbol{\Theta}$ a region of interest within the parameters space (containing the expected values of the parameters).

It is well known \cite{fedorov1972theory} that the number of $b$-values (design points) must be grater (or equal) than the total number of parameters $P$ (4 in this case) in order for $\mathbf{M}$ not to be singular. Moreover, in a \emph{continuous design} \cite{fedorov1972theory} (i.e. when the number of measurements taken at the design points is very large)
the number of design points is limited superiorly by $P(P+1)/2+1$ (11 in the present case).
In passing to a \emph{discrete design} (i.e. a single measure is taken at each design point, as is the case for IVIM studies)
the number could be greater than this limit.
However, we used 20 as maximum because of the previous considerations (see Introduction section) concerning the duration of the exam

\subsection{Comparison of designs}\label{sec:accuracy}
To compare different designs we used the Cramer-Rao lower bound (CRLB) \cite{sansone2015geometrical,smith2005cramerrao,fedorov1972theory,bates1988nonlinear}
Assuming the parameters estimates are not biased, 
according to the Cramer-Rao lower bound theorem \cite{smith2005cramerrao}
the achievable accuracy is given 
 by the diagonal elements of the inverse  Fisher information matrix.
The inverse of the Fisher matrix is given by: 
\begin{equation} 
\mathbf{M}^{-1} = \sigma^2
\left( \begin{array}{cc}
1 & \mathbf{0}\\
\mathbf{0} & S_0^{-1} \mathbf{I} \\
\end{array} 
\right)
(\mathbf{J}^T \mathbf{J})^{-1}
\left( \begin{array}{cc}
1 & \mathbf{0} \\
\mathbf{0} & S_0^{-1} \mathbf{I} \\
\end{array} 
\right)
\end{equation}

Calling $j^{-1}_{kk}$ with $k = 1,\ldots,4$, the diagonal elements of the matrix $(\mathbf{J}^T\mathbf{J})^{-1}$ we have the following bounds for the estimated parameters:
\begin{equation}\label{eq:crlb}
\left( \begin{array}{ccc}
\sigma_{f}^2 & & \\
& \sigma_{D_s}^2 & \\
& & \sigma_{D_f}^2
\end{array} 
\right) \succeq \frac{\sigma^2}{S_0^2}  
\left( \begin{array}{ccc}
j^{-1}_{22} & & \\
& j^{-1}_{33} & \\
& & j^{-1}_{44}
\end{array} 
\right)
\end{equation}
and $\sigma^2 j^{-1}_{11}$ is the CRLB of the variance of $S_0$ estimator, which is of no interest here. The square root of CRLB ($\sigma_f$, $\sigma_{D_s}$,$\sigma_{D_f}$) can be considered the achievable accuracy for the parameter estimate.

We have computed the square root of CRLB per each parameter
at the $\mathbf{b}_{opt}$ D-optimal values (eq. \ref{eq:maxFisher}), 
over the whole parameters region used in the optimisation process.

\subsection{Numerical optimisation}

\subsubsection{Parameters region}
The region of the parameter space $\boldsymbol{\Theta}$ has been chosen 
on the basis of parameters values 
found in published literature:
see table \ref{tab:parameterSpace}.

\begin{table}
\caption{Values of IVIM parameters used in previous simulation studies.
The notation used is Octave-like: the number between two colons (:) is the \textit{increment} 
between the minimum and the maximum of the range.}
\label{tab:parameterSpace}
\centering
\begin{tabular}{ccc}
\hline
Study & $D_s \cdot 10^{-3} [\mathrm{mm^2/s}]$ & $D_f \cdot 10^{-3} [\mathrm{mm^2/s}]$  \\ 
\hline
 \cite{jambor2014optimization} & 0.23, 0.7 & 2.52, 2.9 \\ 
\cite{zhang2013cramer} & 1, 0.5:6 & 2.1, 15.5:81 \\ 
\cite{lemke2011toward} & 1, 1.5 & 11, 16.5, 61 \\ 
this & 0.1:0.1:2, 2:1:6 & 1:1:20, 20:10:90  \\
\hline
\end{tabular}
\end{table}

\subsubsection{Range of candidate $b$ values}
Also the range of $b$ values has been chosen after examination of literature \cite{Koh:2011aa,lemke2011toward}.
In particular, we have used the following set of $b$-values given as in a Octave-like notation $b$ = 0:10:200, 250:50:1000, (37 values).

\subsubsection{Exact design}\label{sec:maxdet}
In principle, according to the max-min criterion (eq. \ref{eq:maxFisher}), 
in order to find the optimal $\mathbf{b}$, it is required to test all the combinations of $b$-values taken from the series of candidate $b$-values. 
Per each combination of $b$-values, the determinant of eq. \ref{eq:maxFisher} must be evaluated on a grid comprising 
all interesting values $(D_s,D_f)$ within the parameter space $\boldsymbol{\Theta}$ and
the minimum value over this region has to be found.
Afterwards, the $\mathbf{b}$ giving  maximum of all minima must be found.
This guarantees  that the Fisher determinant evaluated at $\mathbf{b}_{opt}$ 
is the highest possible over  the entire  parameters region.

However, this approach implies large computational times for designs with 
a number of design points greater than the minimum (i.e. 4, see section \ref{sec:design}): 
this makes also impractical  the design based  over a  large region of the parameters space  
or the use of an large candidate set of $b$-values.
For these reasons, in the next section we propose to use a fast but approximate design.

\subsubsection{Approximate design}\label{sec:algorithmApprox}
As observed the minimum number of design points which gives non singular Fisher matrix for 
the IVIM model is 4.
As a starting point, we found the exact optimal 4 points design using the method described in the previous section : $\mathbf{b}_{opt} (4)  = [0,80,500,1000]$.
In particular, we  evaluated the determinant of the Fisher matrix over all the parameters space $\boldsymbol{\Theta}$
and using as  $\mathcal{B}$ the space of \emph{all}  possible combinations of 4 points chosen among the 37 candidate $b$ values chosen. The resulting number of combinations in this case is $\scalemath{0.75}{
\left( \begin{array}{c}  {37} \\ {4} 
\end{array} \right) }= 66045$.

However, as observed, with increasing number of design points, the number of possible combination increases quickly (e.g. $\scalemath{0.75}{
\left( \begin{array}{c}  {37} \\ {5} \end{array} \right)} = 435897$),
and the computational time increases as well.
In order to limit the computational time we proceeded as follows.

Let $\mathbf{b}(N) = [b_1, \ldots,b_N]$ a design set with $N$ points. 
Starting from an optimal design $\mathbf{b}(N)$ we can construct a new design 
$\mathbf{b}(N+1) = [\mathbf{b}(N),~b_{new}]$
adding a new $b_{new}$ value.
The matrix $\mathbf{H}$ corresponding to this new set of design points is (dropping down the dependence on $D_s$ and $D_f$ for simplicity):
\begin{equation}
\mathbf{H}^T (\mathbf{b}(N+1)) = [\mathbf{H}^T (\mathbf{b}(N)), ~ \mathbf{r}(b_{new})] 
\end{equation}
where 
\begin{equation}
\mathbf{r}(b_{new}) =  \left( 
\begin{array}{c}
\exp(-b_{new} D_s) \\ 
\exp(-b_{new} D_f ) \\
 b_{new}\exp(-b_{new} D_s) \\
  b_{new}\exp(-b_{new} D_f)
\end{array} \right)
\end{equation}

As we are interested in calculation of the determinant of the Fisher matrix we observe that \cite{fedorov1972theory} (we drop the dependence from $b_{new}$,$D_s$ and $D_f$ in order to simplify notation) :
\begin{eqnarray}\label{eq:increment}
& & || \mathbf{M}(\mathbf{b}(N+1)) || = \\
\nonumber  & & = || \mathbf{M}(\mathbf{b}(N))||
 \cdot (1 + \mathbf{r}^T (\mathbf{H}^T(N) \mathbf{H}(N) )^{-1}  \mathbf{r} )
\end{eqnarray}

In order for the new design being optimal this determinant must be maximum. This implies that $b_{new}$ must maximise equation \ref{eq:increment} over the parameters region. Using equation \ref{eq:increment} and starting from the optimal design $\mathbf{b}(4)$,
designs with a fixed $N$ can be constructed iteratively, evaluating one only matrix inverse and performing $\sim N$ computations 
instead of examining all the possible combinations.

Unfortunately, this algorithm does not provide an {absolutely} optimal design, because there can be different combinations of $N+1$ points extracted from the original set, giving a higher determinant.
This algorithm provides therefore a \emph{sub-optimal} solution. However, in our experience (we tested the case $N=5$ and $N=6$), the accuracy attainable with the sub-optimal solution is not far away from the optimal solution (see figure \ref{fig:comparison6pointsDesign}).

\section{Results}

All optimisations have been performed in Octave \cite{octave}.

Table \ref{tab:optimalExact} reports the exact designs obtained for $N = 4, 5, 6$ respectively.
They have been obtained considering all the possible combinations of $N$ $b$-values chosen from the candidate set.

Table \ref{tab:optimalApproximate} reports the approximate designs obtained using the algorithm described in section \ref{sec:algorithmApprox}. The first four entries of the table coincide with the 4 points design; starting from this one  should add the successive entries in order to obtain designs with more than 4 points. For example a 6 points design includes $0, 80, 500, 1000, 190, 600$.

\begin{figure*}[!t]
\centerline
{
\subfigure[ exact, f]{\includegraphics[width=0.45\textwidth]{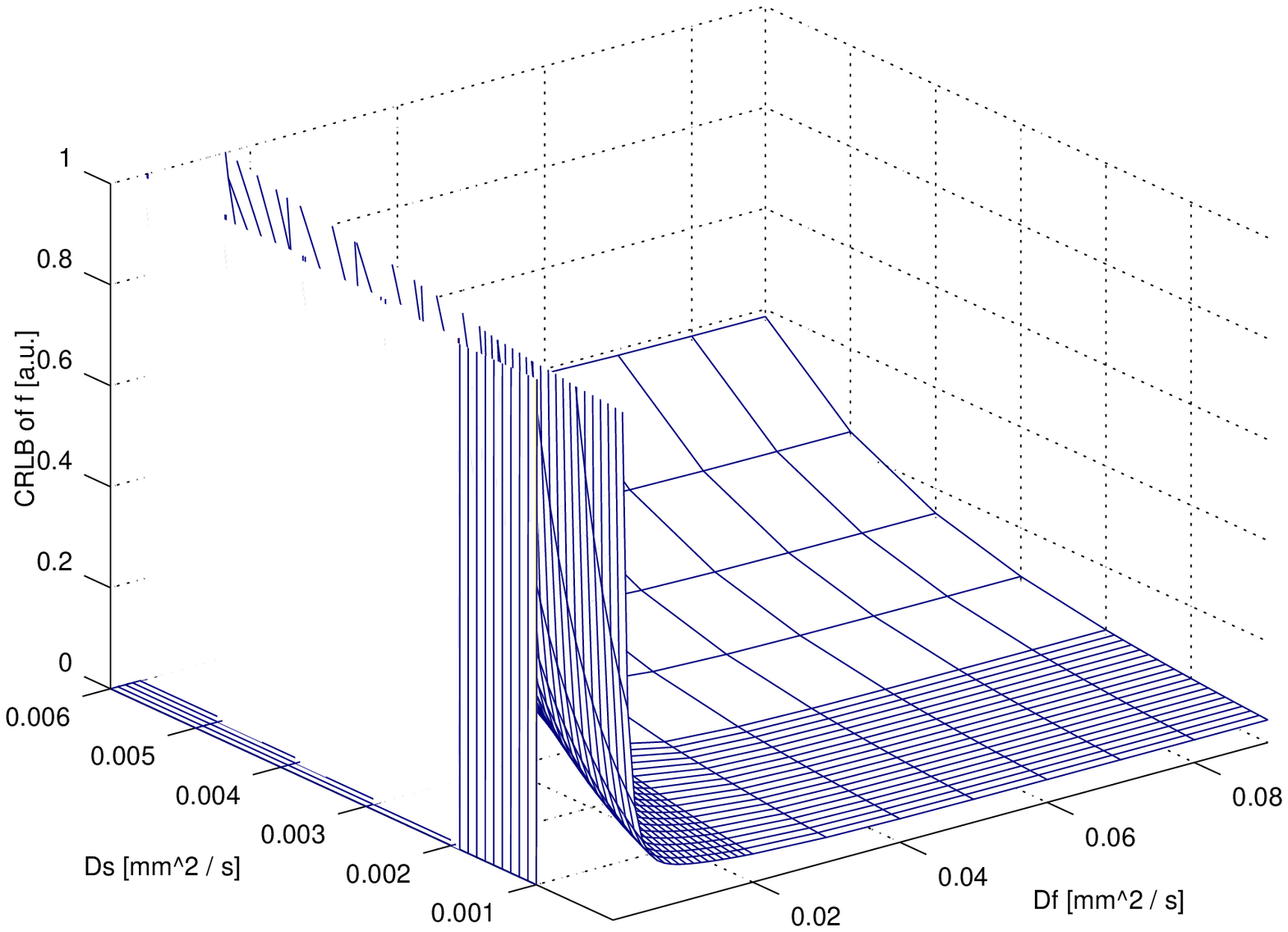} }
\label{fig:f6exact}
\subfigure[ approx, f]{\includegraphics[width=0.45\textwidth]{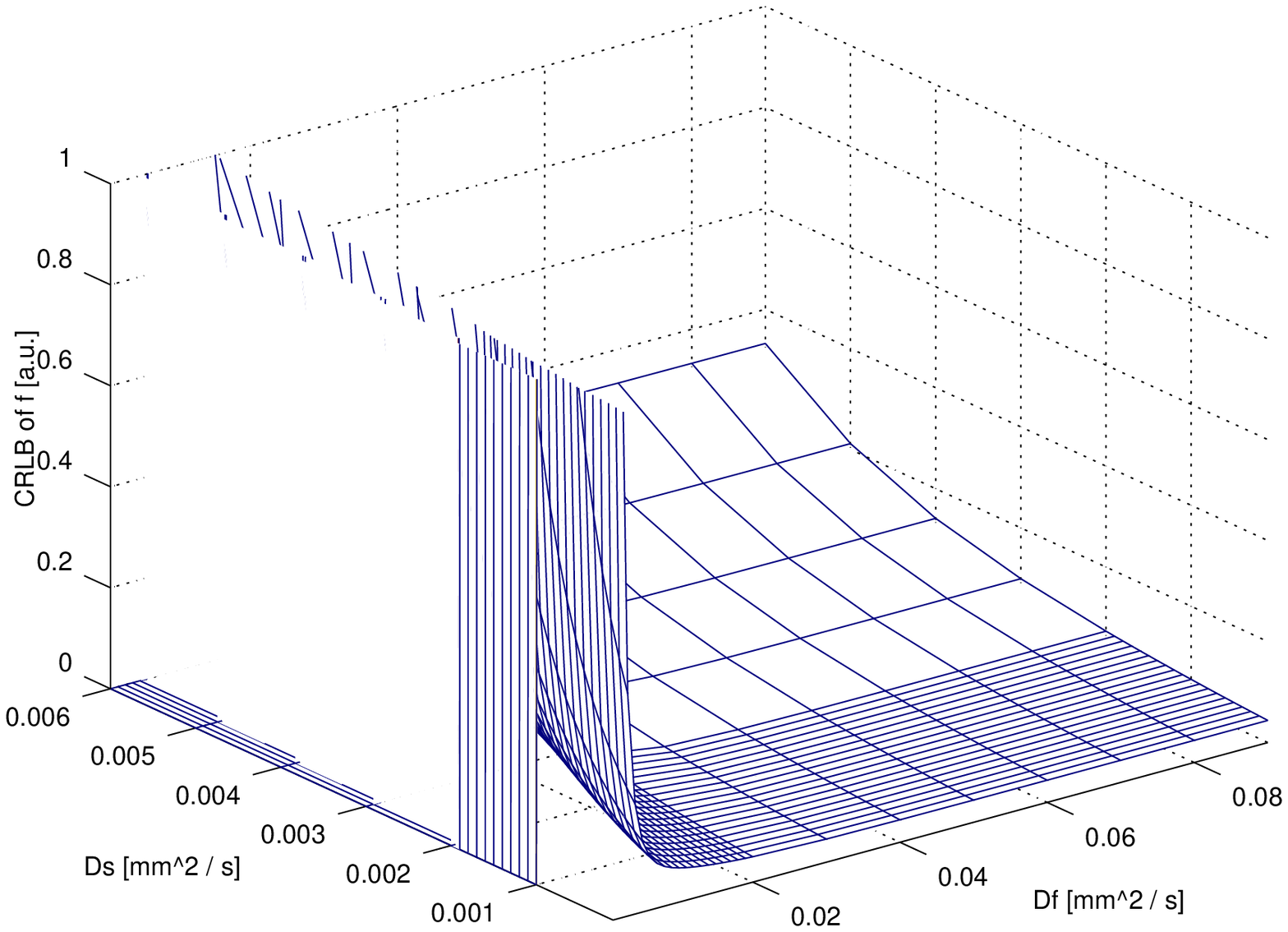} }
\label{f6approx}
}
\centerline
{
\subfigure[ exact, $D_s$]{ \includegraphics[width=0.45\textwidth]{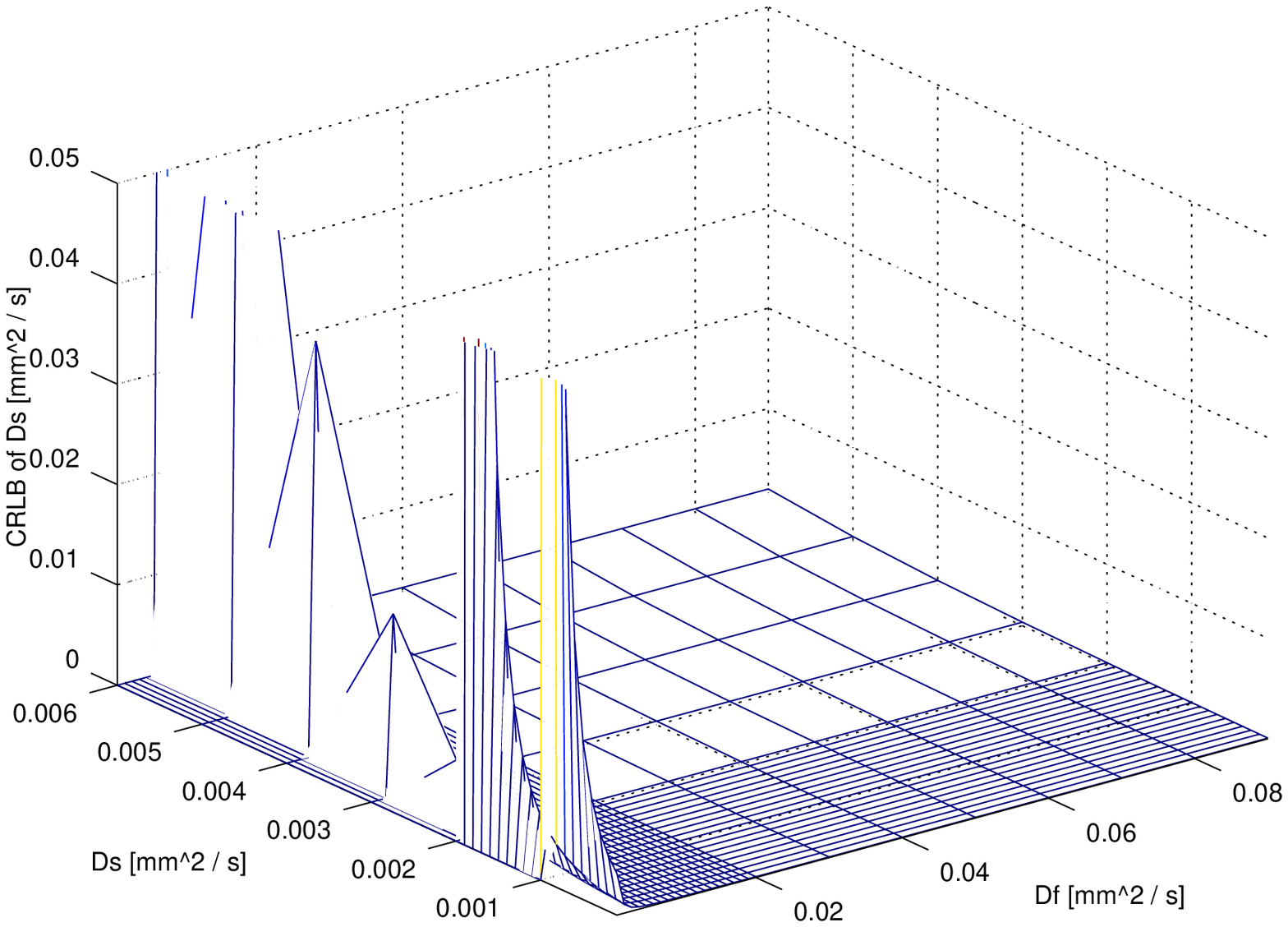} }
\label{fig:ds6exact}
\subfigure[approx $D_s$]{ \includegraphics[width=0.45\textwidth]{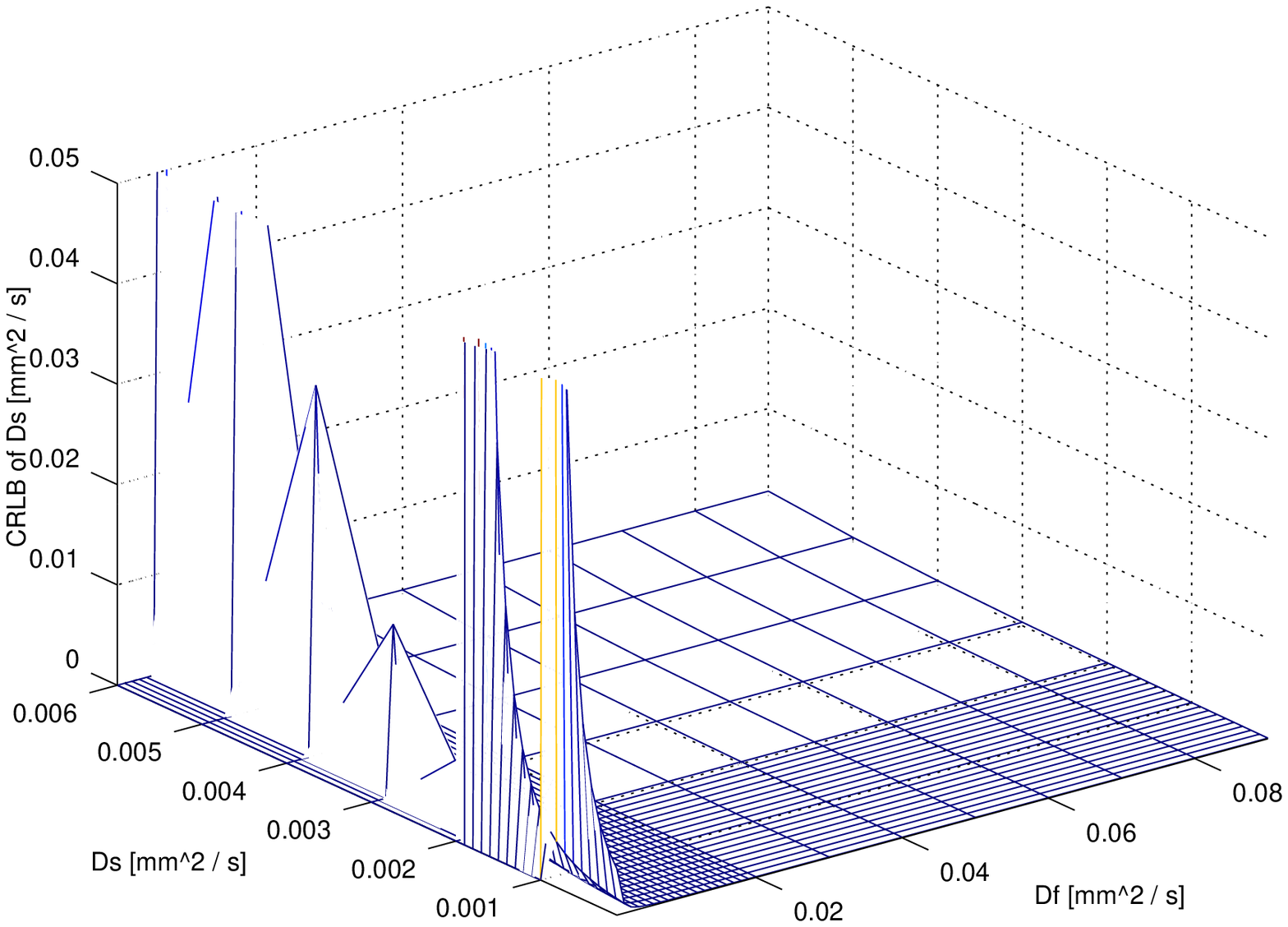} }
\label{fig:ds6approx}
}
\centerline
{
\subfigure[ exact, $D_f$]{ \includegraphics[width=0.45\textwidth]{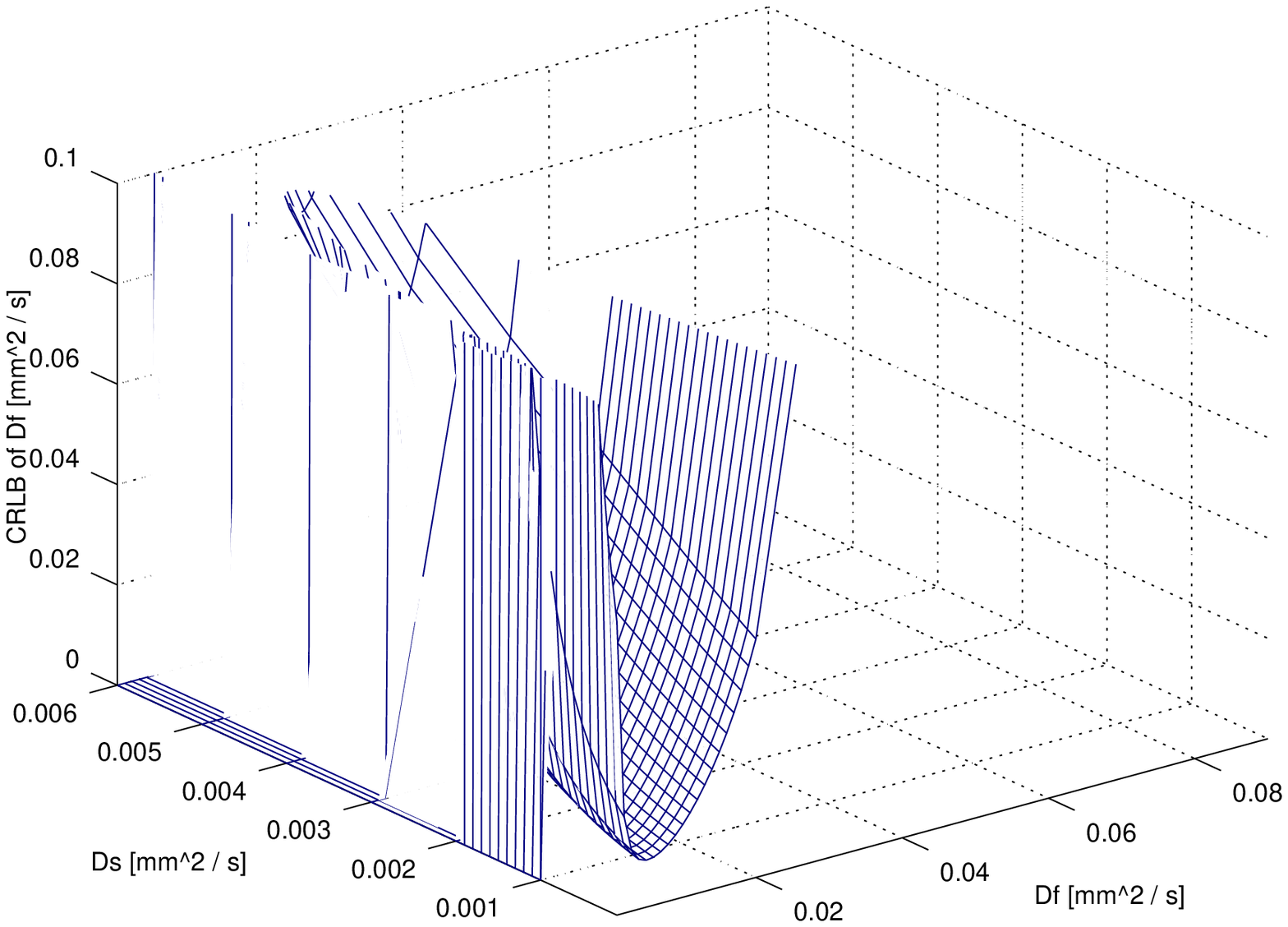} }
\label{fig:df6exact}
\subfigure[approx, $D_f$ ]{ \includegraphics[width=0.45\textwidth]{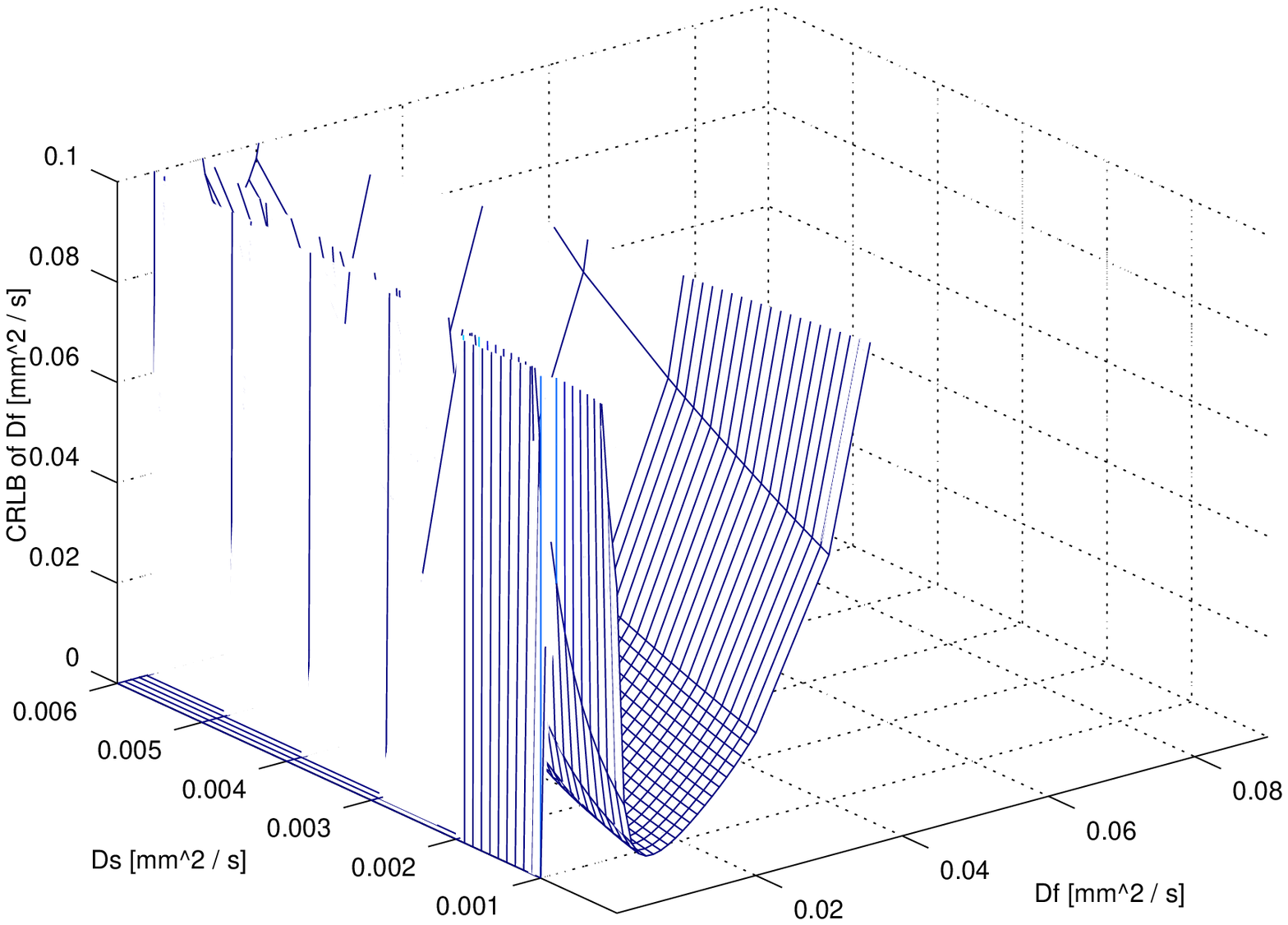} }
\label{fig:df6approx}
}
\caption{Comparison between the exact 6 points design and the approximate one.
 Cramer -Rao lower bound (CRLB) is reported for each parameter for both designs.
 For all figures were used the following settings:  $f = 0.1$, $S_0 = 100$, $\sigma = 3$ corresponding to a SNR of $S_0/\sigma \approx 33.3$ (see section \ref{sec:noise}).
 Left and right columns correspond to exact and approximate design respectively.
 Top, middle and bottom rows correspond to the parameters $f$, $D_s$ and $D_f$ respectively.
 }
\label{fig:comparison6pointsDesign}
 \end{figure*}

In order to compare  exact designs with approximate ones the CRLB can be used:
in particular, for illustrative purposes, the comparison between the exact 6 points design and  the approximate one is reported in figure \ref{fig:comparison6pointsDesign}. The attainable accuracy (CRLB) has been computed over the whole parameter region. It is clear that per each parameter the CRLB with exact design is not very different from the corresponding accuracy attainable using the approximate design.

\begin{figure}[!t]
 \centering
{\includegraphics[width=0.45\textwidth]{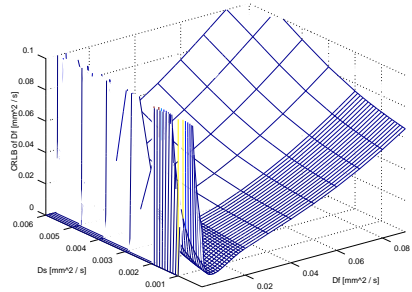} }
 \caption{CRLB for the $D_f$ parameter using a 20 points design. This figure must be compared with the bottom row in fig. \ref{fig:comparison11pointsDesignPerfect}.
 }
 \label{fig:20points}
 \end{figure}

\begin{figure*}[!t]
\centerline
{
\subfigure[ approx, f]{\includegraphics[width=0.45\textwidth]{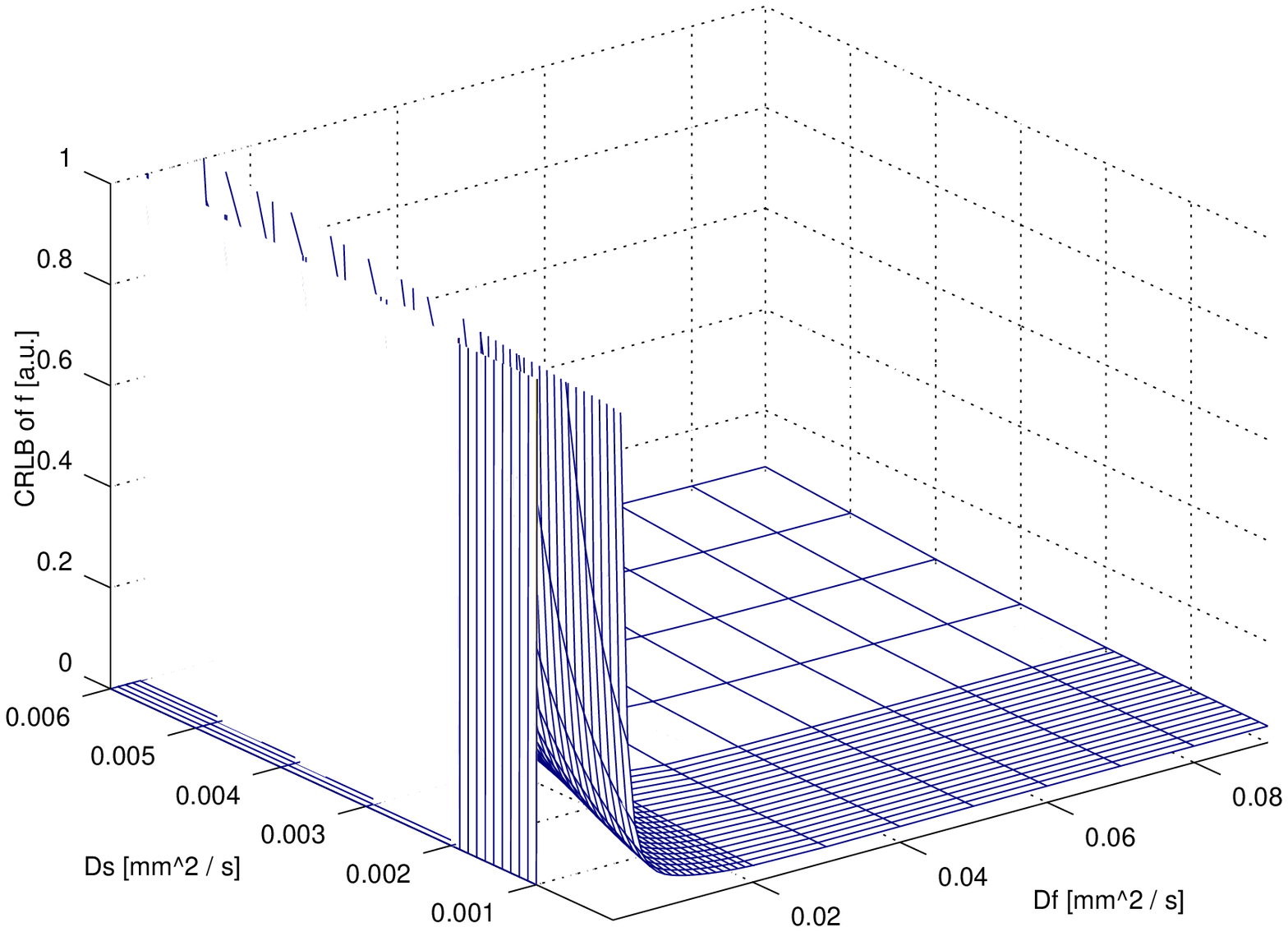} }
\label{fig:f11approx}
\subfigure[ ideal, f]{\includegraphics[width=0.45\textwidth]{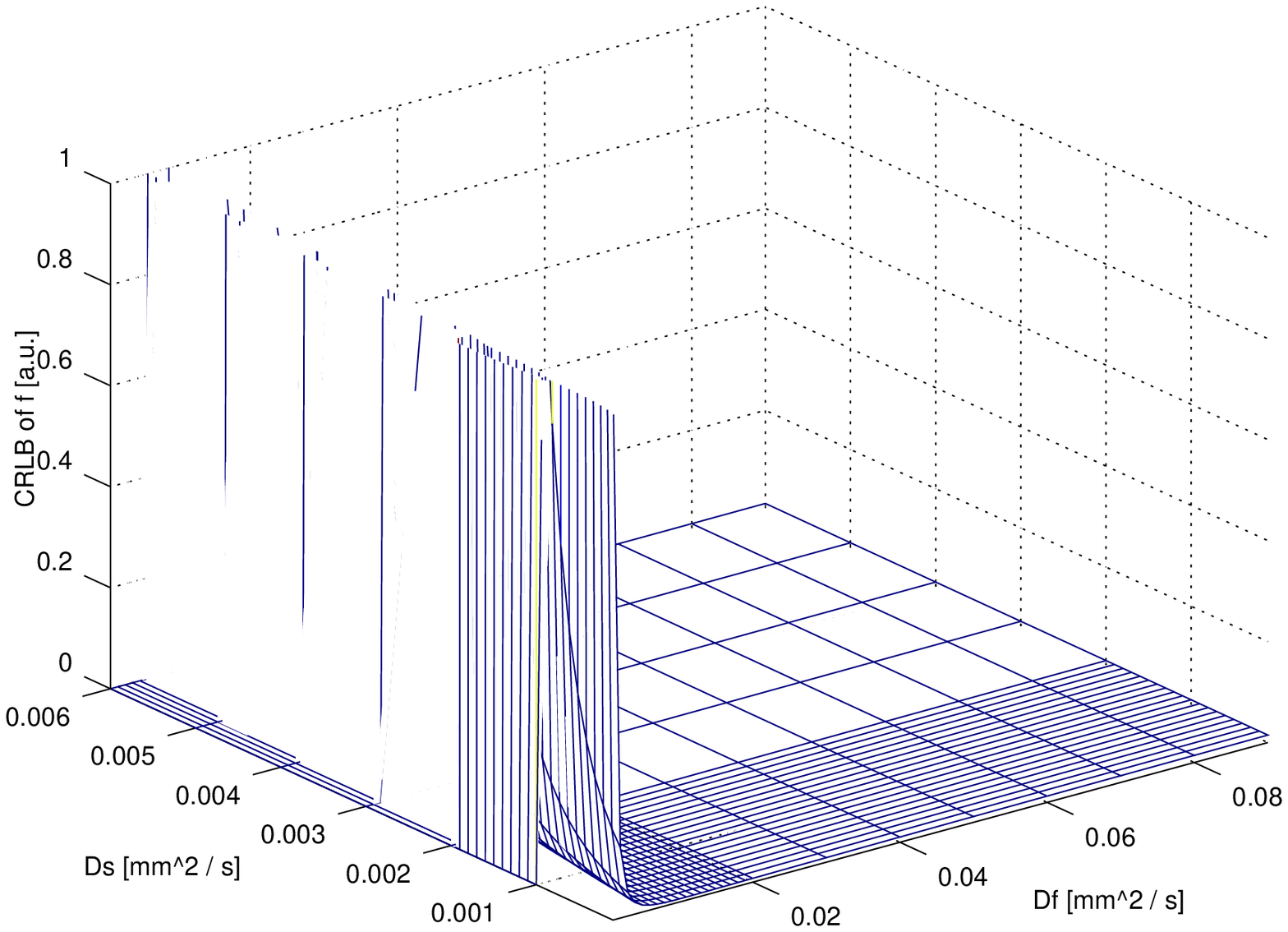} }
\label{fig:fperfect}
}
\centerline
{
\subfigure[ approx, $D_s$]{ \includegraphics[width=0.45\textwidth]{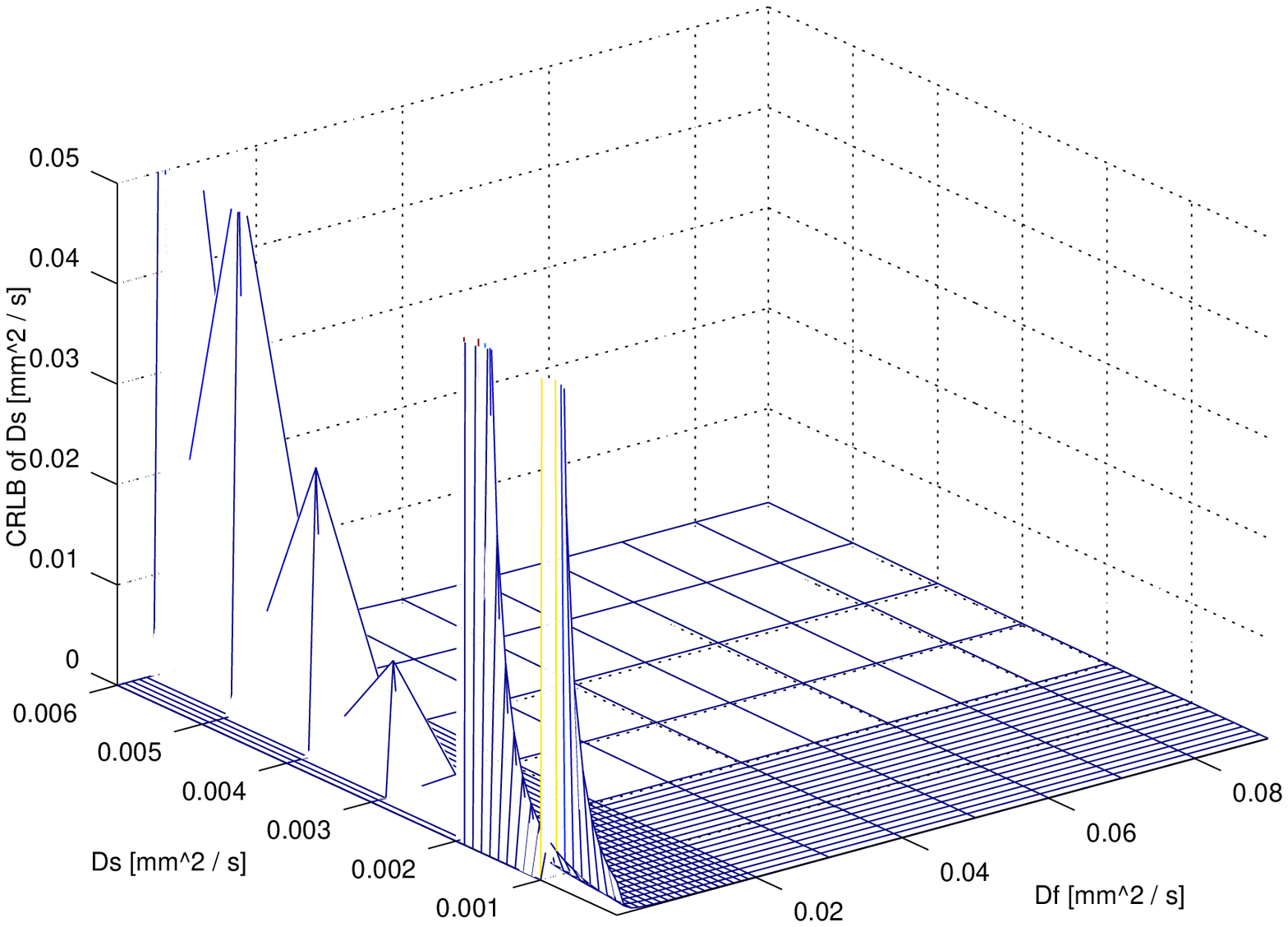} }
\label{fig:ds11approx}
\subfigure[ideal $D_s$]{ \includegraphics[width=0.45\textwidth]{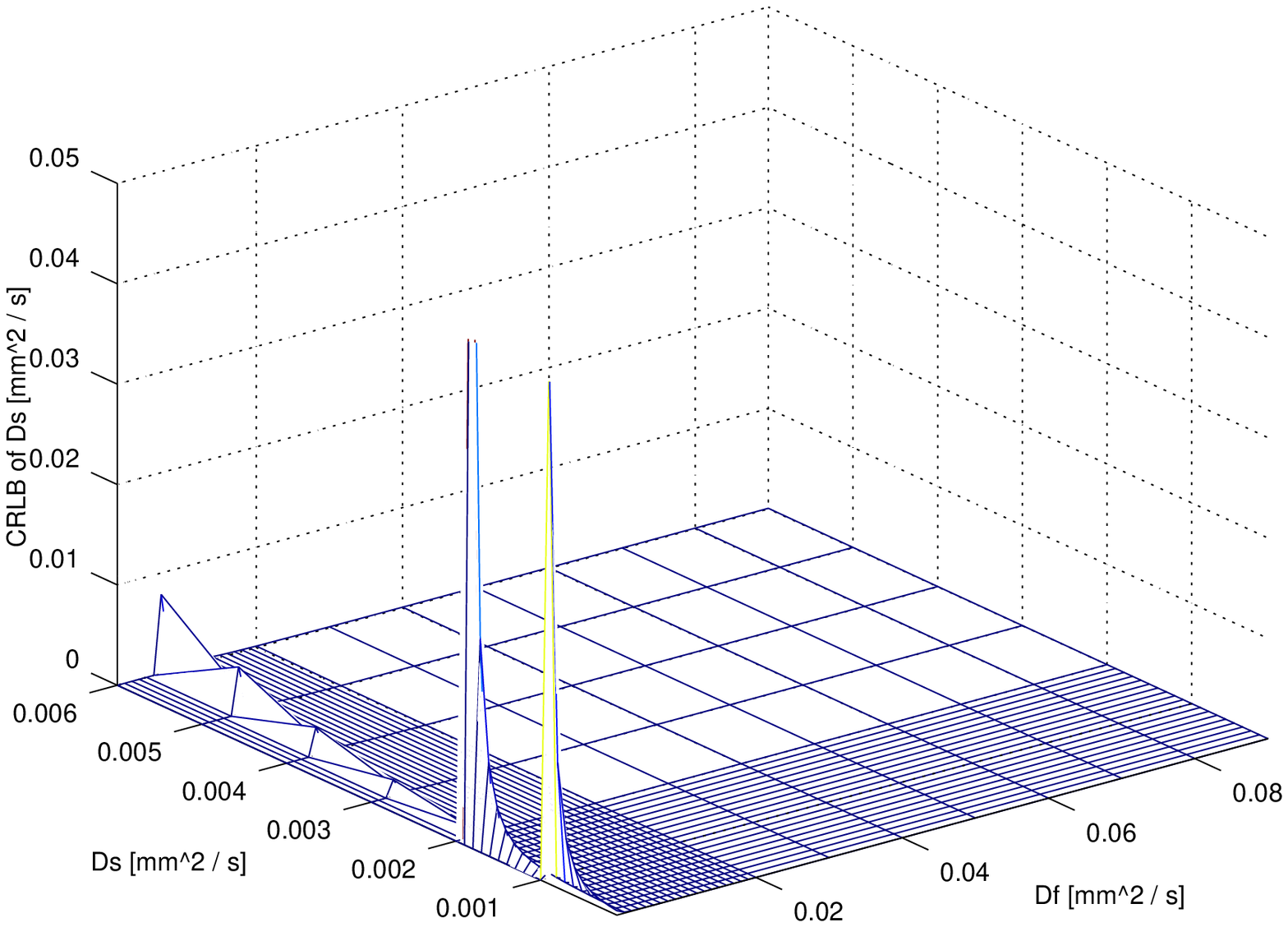} }
\label{fig:dsperfect}
}
\centerline
{
\subfigure[ approx, $D_f$]{ \includegraphics[width=0.45\textwidth]{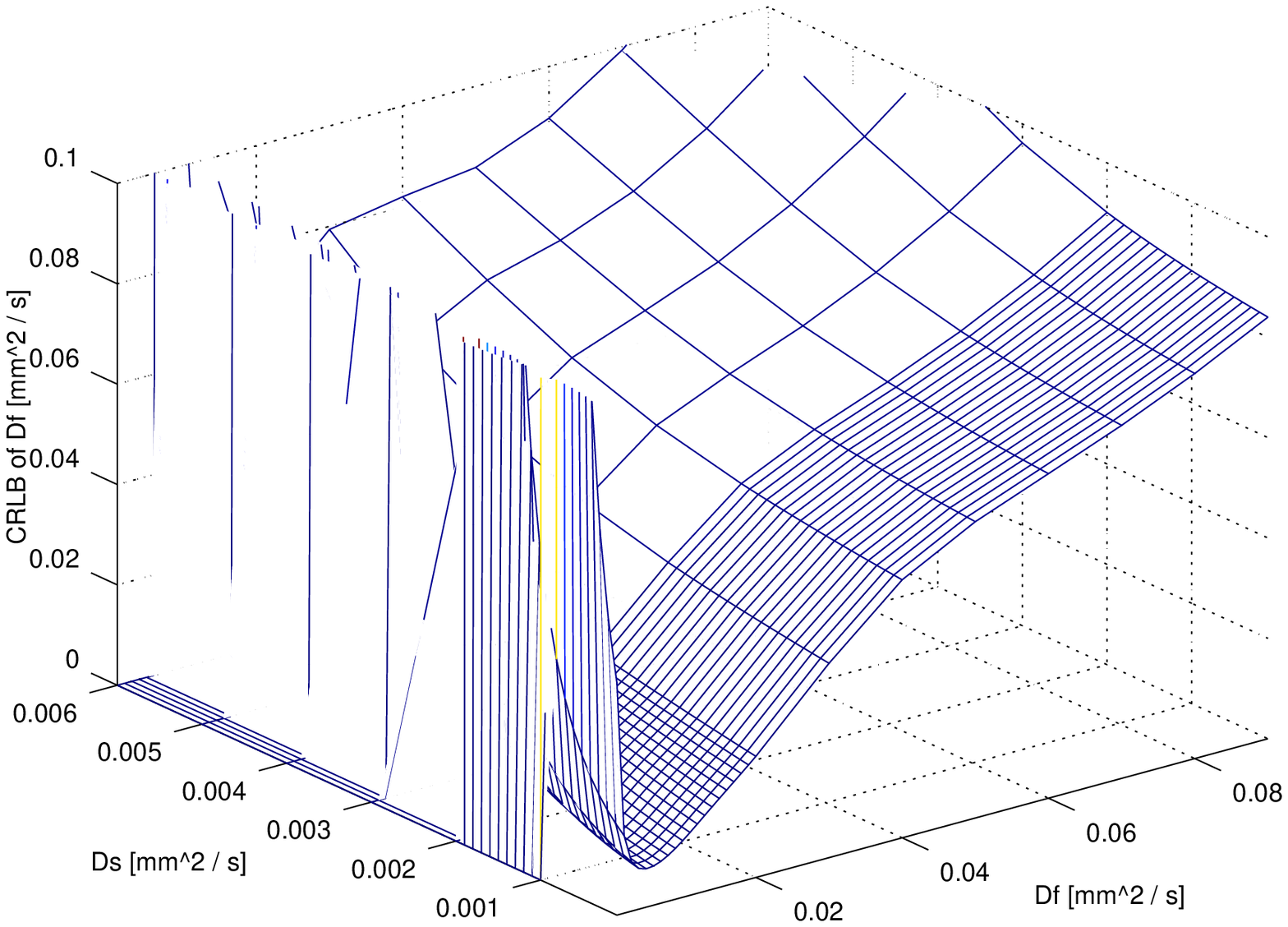} }
\label{fig:df11approx}
\subfigure[ideal, $D_f$ ]{ \includegraphics[width=0.45\textwidth]{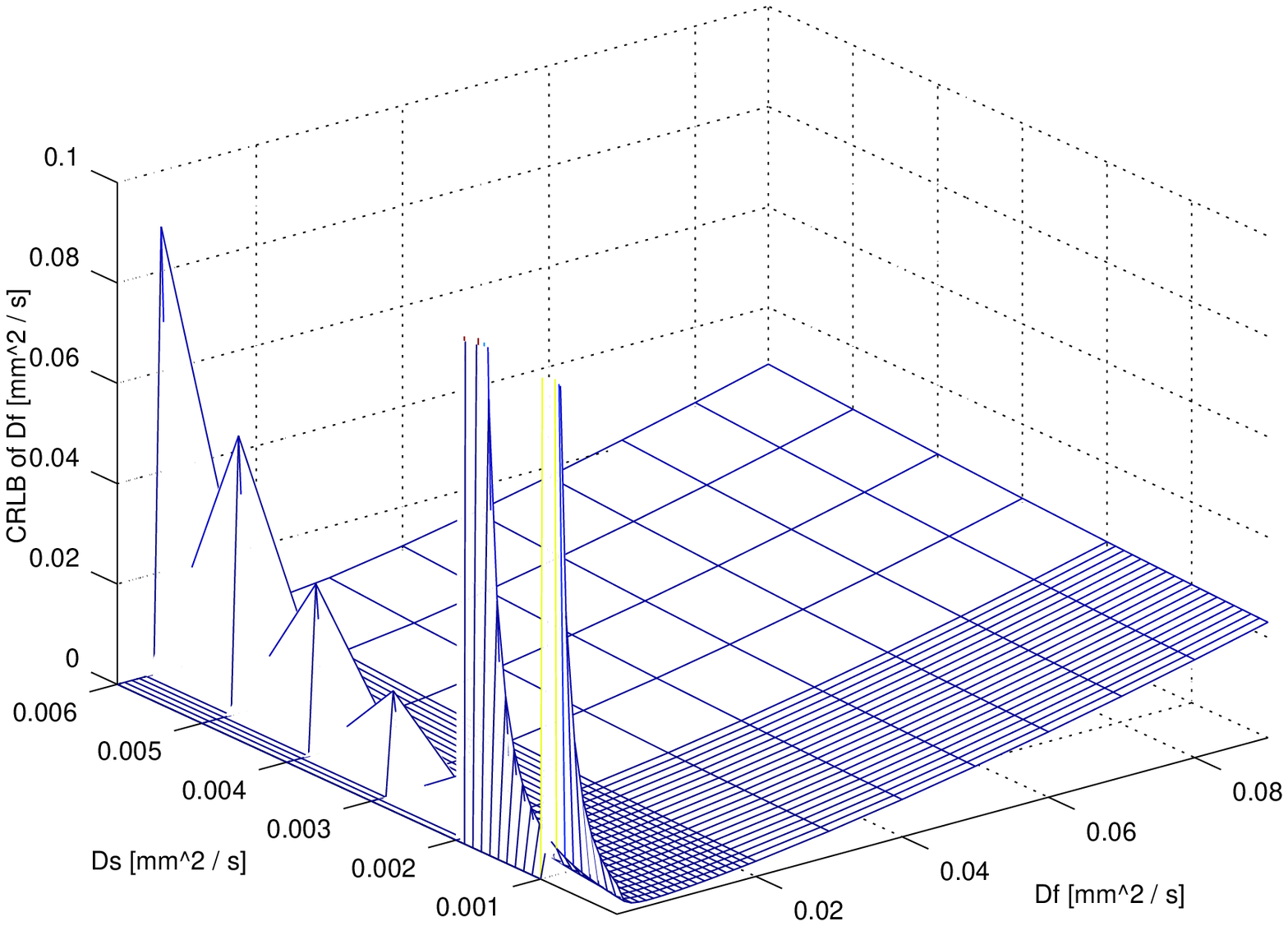} }
\label{fig:dfperfect}
}
\caption{Comparison between the approximate 11 points design and the `ideal' one.
 Cramer-Rao lower bound (CRLB) is reported for each parameter for both designs.
 For all figures were used the following settings:  $f = 0.1$, $S_0 = 100$, $\sigma = 3$ corresponding to a SNR of $S_0/\sigma \approx 33.3$ (see section \ref{sec:noise}).
 Left and right columns correspond to approximate and `ideal' design respectively.
 Top, middle and bottom rows correspond to the parameters $f$, $D_s$ and $D_f$ respectively.
 }
\label{fig:comparison11pointsDesignPerfect}
 \end{figure*}

It is useful to compare the accuracy attainable  with design using a small number of points (e.g. 11) with respect to the higher accuracy attainable using a very large $N=1000$ design (e.g. $b = 0:1:1000$, \emph{ideal} design).
The comparison is reported in figure \ref{fig:comparison11pointsDesignPerfect}.
It is evident that no large difference is revealed for $f$ and $D_s$; on the contrary, `ideal' design gains a significant improvements with respect to the 11 points design: however, in a clinical scenario this might not justify the extra time required.
Further insight on this issue can be achieved observing fig \ref{fig:20points}.
As a matter of fact, the use of 20 points does not improve with respect to 11 points design.

In figures \ref{fig:comparison6pointsDesign} and \ref{fig:comparison11pointsDesignPerfect}
a value of $f= 0.1$ has been used; other parameters were $S_0 = 100$ and $\sigma = 3$ corresponding to an SNR of $S_0/\sigma \approx 3.33$. These values can be considered to be representative of real values from published literature.

\begin{figure}[!t]
 \centering
 \includegraphics[width=0.45\textwidth]{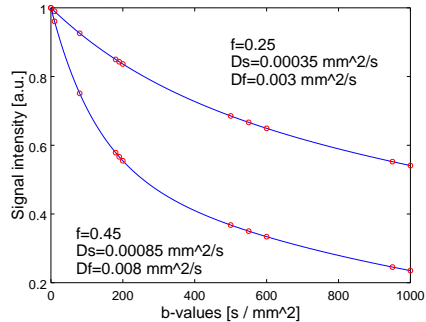} 
 \caption{Example of normalised ($S(b)/S_0$) IVIM curve with various parameters. The solid line represents the theoretical curve; circles indicate the $b$-values sampled using the results of the present study.
 }
 \label{fig:ivimExample}
 \end{figure}

In order to give an idea of the location of the design points, figure \ref{fig:ivimExample} shows an example of DW data vs $b$ for two sets of parameters with the 11 design points superimposed.

\begin{table}[!t]
\renewcommand{\arraystretch}{1.3}
\caption{Optimal designs calculated using the exact algorithm described in section \ref{sec:design}. 
Per each number of design points ($N$) the combination of $b$-values chosen from the candidate set is reported.}
\label{tab:optimalExact}
\centering
\begin{tabular}{ccccccc}
\hline
$N$ & \multicolumn{6}{c}{Exact Design} \\
\hline
4 & 0 & 80 &   500  & 1000 & & \\
5  & 0 & 120 & 200 & 550 &   1000 & \\
6 & 0 & 120 & 200 & 550 & 600 & 1000 \\
\hline
\end{tabular}
\end{table}

\begin{table}[!t]
\renewcommand{\arraystretch}{1.3}
\caption{Approximate optimal design. Starting from the exact 4 points design (the first 4 entries in the table) we derived  additional $b$-values iteratively using the algorithm proposed in section \ref{sec:design}. We report results up to $N=20$ $b$-values.
}
\label{tab:optimalApproximate}
\centering
\begin{tabular}{ccccccccccc}
\hline
0 &  80 &  500 & 1000 & 190 & 600 & 950 \\
 10 & 180 & 550 & 200  & 900 & 20 & 650 \\
  170 & 450 &  160 & 30 &  850 &  400 & \\
\hline
\end{tabular}
\end{table}

\section{Discussion}
The aim of this paper was to design the $b$-values
for diffusion weighted MRI
in an optimal manner suitable for accurate estimation
of intra-voxel incoherent motion parameters
according to the model in eq. \ref{eq:ivim2}.

The design has been conducted according to the principles of the D-optimal approach.
Optimal combination of $b$-values
has been chosen within a set of predefined values
taken from the literature.
As the design is affected from the parameters $D_s,D_f$
an exhaustive optimisation within a predefined  parameter region 
has been made.

Our results are in line with other studies in literature 
\cite{jambor2014optimization,lemke2011toward}, 
which have been conducted via Monte Carlo simulation.
Two main disadvantages of the approach via Monte Carlo simulation
are: first, in order to have statistical accuracy a large number of simulation must be performed (typically 1000) and per each simulated noisy  curve estimation of parameters must be performed (e.g. via least squares fitting); second, due to computational time only a small portion of the parameters region can be explored in reasonable time.
In fact, least squares fitting of noisy simulated curves
is time consuming  and if $S_0$ is neglected it 
might be not suitable to the noise structure on the data (see section \ref{sec:noise}).
Finally, as the parameters values ($f,D_s,D_f$) are not known prior to the MR exam
it would be desirable to have a set of $b$ values  
optimised over a large portion of the parameter space.

The approach followed in this study
could overcome the above mentioned disadvantages of the Monte Carlo approach.
In particular, we showed (section \ref{sec:design}) that the search for a D-optimal design
can be addressed in the 2D space of the diffusion parameters only $(D_s,D_f)$
without considering $S_0$ and $f$: this dramatically reduces the computational load.

Moreover, we propose a fast algorithm for finding an approximate design: 
on the basis of CRLB analysis we showed that the approximate design is comparable to the exact design
(at least in the case of 5 and 6 points designs).
In fact, inspection of table \ref{tab:optimalApproximate} and \ref{tab:optimalExact} reveals that 
the approximate designs with 5 and 6 points are only slightly different from the exact 5 and 6 points design.  
These similarities is confirmed analysing the CRLB in figure \ref{fig:comparison6pointsDesign}.
From these similarities we infer that also for higher values of $N$
the exact and approximate designs might be very similar (for $N \rightarrow \infty$ the exact and approximate design should converge \cite{fedorov1972theory}).

The comparison of 11 points design with the 1000 points design (see figure \ref{fig:comparison11pointsDesignPerfect}) suggest that even with an ideal design
the uncertainty over $D_f$ is limited. 
Furthermore, use of 20 points only slightly improves the accuracy with respect of a 11 points design:
this is a useful information in a clinical setting.

A consideration about the hypothesis underlying  this study must be made.
We used the approximation of Gaussian noise
with mean $\approx S$ instead of $\sqrt{S^2 + \sigma^2}$ 
because our aim was the design of $b$-values and not the estimation
of parameters. As a matter fact, the estimation of parameters is slightly 
biased and care must be taken in the choice of estimator  \cite{kristoffersen2007optimal}

One final remark is the following.
Our study has shown that it is possible to search for an optimal $N+1$-point design
starting from an (approximately) optimal design of $N$ points.
This procedure is very fast.
This might suggest an \emph{adaptive} strategy for D-optimal design
which could be performed during clinical exam:
after a first scan using the minimum set of 4 $B$-values,
an estimate $\boldsymbol{\theta}_0$ of the parameters is performed over the region of interest (within the image), 
opportunely selected by the radiologist;
a spatial average of the first estimate
might be used for the design of the 5th point;
the procedure can be repeated giving estimates $\boldsymbol{\theta}_1$, $\boldsymbol{\theta}_2$ and so on. The radiologist can stop after reaching the desired accuracy in the estimates or after a reasonable time.
This adaptive procedure has not been investigated here but will be subject of future studies.

\section{Conclusion}
The design of the $b$-values for optimal estimation of IVIM parameters 
can be addressed using a D-optimal strategy.
In this study we have shown that the optimal design does not depend on 
perfusion fraction $f$ and therefore the search can be performed in a 2-D space $(D_S,D_f)$;
moreover, as an exact exhaustive search is still time consuming,
 we have proposed an iterative algorithm for searching an approximate design  starting from an optimal design formed by 4 points.


%



\section*{Acknowledgment}
The authors would like to thank Dr. Augusto Aubry at
the Dept. of Electrical Engineering and Information Technologies of 
University of Naples `Federico II'
for fruitful discussions.


\begin{thebibliography}{10}
\providecommand{\url}[1]{#1}
\csname url@samestyle\endcsname
\providecommand{\newblock}{\relax}
\providecommand{\bibinfo}[2]{#2}
\providecommand{\BIBentrySTDinterwordspacing}{\spaceskip=0pt\relax}
\providecommand{\BIBentryALTinterwordstretchfactor}{4}
\providecommand{\BIBentryALTinterwordspacing}{\spaceskip=\fontdimen2\font plus
\BIBentryALTinterwordstretchfactor\fontdimen3\font minus
  \fontdimen4\font\relax}
\providecommand{\BIBforeignlanguage}[2]{{%
\expandafter\ifx\csname l@#1\endcsname\relax
\typeout{** WARNING: IEEEtran.bst: No hyphenation pattern has been}%
\typeout{** loaded for the language `#1'. Using the pattern for}%
\typeout{** the default language instead.}%
\else
\language=\csname l@#1\endcsname
\fi
#2}}
\providecommand{\BIBdecl}{\relax}
\BIBdecl

\bibitem{Koh:2011aa}
D.-M. Koh, D.~J. Collins, and M.~R. Orton, ``Intravoxel incoherent motion in
  body diffusion-weighted mri: reality and challenges,'' \emph{AJR Am J
  Roentgenol}, vol. 196, no.~6, pp. 1351--61, Jun 2011.

\bibitem{zhang2013cramer}
Q.~Zhang, Y.-X. Wang, H.~T. Ma, and J.~Yuan, ``Cramer-rao bound for intravoxel
  incoherent motion diffusion weighted imaging fitting,'' in \emph{Engineering
  in Medicine and Biology Society (EMBC), 2013 35th Annual International
  Conference of the IEEE}.\hskip 1em plus 0.5em minus 0.4em\relax IEEE, 2013,
  pp. 511--514.

\bibitem{lemke2011toward}
A.~Lemke, B.~Stieltjes, L.~R. Schad, and F.~B. Laun, ``Toward an optimal
  distribution of b values for intravoxel incoherent motion imaging,''
  \emph{Magnetic resonance imaging}, vol.~29, no.~6, pp. 766--776, 2011.

\bibitem{jambor2014optimization}
I.~Jambor, H.~Merisaari, H.~J. Aronen, J.~J{\"a}rvinen, J.~Saunavaara,
  T.~Kauko, R.~Borra, and M.~Pesola, ``Optimization of b-value distribution for
  biexponential diffusion-weighted mr imaging of normal prostate,''
  \emph{Journal of Magnetic Resonance Imaging}, vol.~39, no.~5, pp. 1213--1222,
  2014.

\bibitem{le1988separation}
D.~Le~Bihan, E.~Breton, D.~Lallemand, M.~Aubin, J.~Vignaud, and
  M.~Laval-Jeantet, ``Separation of diffusion and perfusion in intravoxel
  incoherent motion mr imaging.'' \emph{Radiology}, vol. 168, no.~2, pp.
  497--505, 1988.

\bibitem{fedorov1972theory}
V.~V. Fedorov, \emph{Theory of optimal experiments}.\hskip 1em plus 0.5em minus
  0.4em\relax Elsevier, 1972.

\bibitem{bates1988nonlinear}
D.~M. Bates and D.~G. Watts, \emph{Nonlinear regression: iterative estimation
  and linear approximations}.\hskip 1em plus 0.5em minus 0.4em\relax Wiley
  Online Library, 1988.

\bibitem{gudbjartsson1995rician}
H.~Gudbjartsson and S.~Patz, ``The rician distribution of noisy mri data,''
  \emph{Magnetic resonance in medicine}, vol.~34, no.~6, pp. 910--914, 1995.

\bibitem{kristoffersen2007optimal}
A.~Kristoffersen, ``Optimal estimation of the diffusion coefficient from
  non-averaged and averaged noisy magnitude data,'' \emph{Journal of Magnetic
  Resonance}, vol. 187, no.~2, pp. 293--305, 2007.

\bibitem{fusco2015use}
R.~Fusco, M.~Sansone, and A.~Petrillo, ``The use of the levenberg--marquardt
  and variable projection curve-fitting algorithm in intravoxel incoherent
  motion method for dw-mri data analysis,'' \emph{Applied Magnetic Resonance},
  vol.~46, no.~5, pp. 551--558, 2015.

\bibitem{probability1984random}
A.~P. Probability, ``Random variables and stochastic processes,'' \emph{McGrow,
  Hill Series Elastical Eng, NY}, 1984.

\bibitem{sansone2015geometrical}
M.~Sansone, R.~Fusco, and A.~Petrillo, ``A geometrical perspective on the 3tp
  method in dce-mri,'' \emph{Biomedical Signal Processing and Control},
  vol.~16, pp. 32--39, 2015.

\bibitem{smith2005cramerrao}
S.~Smith, ``Covariance, subspace, and intrinsic cramer-rao bounds,''
  \emph{Signal Processing, IEEE Transactions on}, vol.~53, no.~5, pp.
  1610--1630, May 2005.

\end{thebibliography}
\end{document}